\documentclass[aps,pre,twocolumn,groupedaddress,showpacs]{revtex4-1}
\newcommand{\bec}[1]{\mbox{\boldmath $ #1$}}
\usepackage{graphicx}
\usepackage{bm}
\begin{document}

\title{Three-dimensional slow Rossby waves in rotating spherical density
stratified convection}
\author{T. Elperin}
\email{elperin@bgu.ac.il}
\homepage{http://www.bgu.ac.il/me/staff/tov}
\author{N. Kleeorin}
\email{nat@bgu.ac.il}
\author{I.~Rogachevskii}
\email{gary@bgu.ac.il}
\homepage{http://www.bgu.ac.il/~gary}
\affiliation{The Pearlstone Center for Aeronautical Engineering
Studies, Department of Mechanical Engineering,
Ben-Gurion University of the Negev, P.O.Box 653, Beer-Sheva 84105,
Israel}
\date{\today}
\begin{abstract}
We develop a theory of three-dimensional slow Rossby waves in rotating
spherical density stratified convection.
The Rossby waves with frequencies which are much smaller than
the rotating frequency, are excited by a non-axisymmetric instability
from the equilibrium based on the developed convection.
These waves interact with the inertial waves
and the density stratified convection.
The density stratification is taken into account using the anelastic
approximation for very low-Mach-number flows.
We study long-term planetary Rossby waves with periods which
are larger than two years.
We suggest that these waves are related to the Southern
Oscillation and El Ni\~{n}o.
The El Ni\~{n}o is an irregularly periodical variation in winds and
sea surface temperatures over the tropical Pacific Ocean, while
the Southern Oscillation is an oscillation in surface air pressure between the tropical eastern and the western Pacific Ocean.
The strength of the Southern Oscillation is characterised by the Southern Oscillation
Index (SOI). The developed theory is applied for the interpretation of the observed periods of the SOI. This study demonstrates a good agreement between the theoretical predictions and the observations.
\end{abstract}

\maketitle

\section{Introduction}

Rossby waves \cite{R39,R40}
which are caused by the combined effect of rotation and curvature of
the surface, exist in various hydrodynamic
flows (see \cite{P78,Z07,CNQ15},
and references therein).
Rossby waves have been found in different geophysical phenomena
(see \cite{P78,Z07,G82,V06,S87,M79},
and references therein), e.g., they are observed in the atmosphere
as the large meanders of the mid-latitude jet stream that
are responsible for the prevailing seasonal weather patterns
and their day-to-day variations (see \cite{P78,G82,V06},
and references therein).
Mesoscale variability in ocean, in scales of tens to hundreds
of km and tens to hundreds of days, occurs as linear Rossby waves
and as nonlinear vortices or eddies (see \cite{CC01,QC03,CS07,CI07,H09,CS11},
and references therein).

Nonlinear dynamics of Rossby waves can cause the formation
of zonal flows or zonal jets (see \cite{N69,R75,CNQ15,DII05},
and references therein).
Zonal flows are coherent structures with a jet-like velocity profile
in anisotropic flows due to the large aspect ratio and planetary rotation.
Zonal flows are produced near the tropopause in the atmosphere
(the polar and subtropical jet streams) and in the oceans
(the Antarctic circumpolar current) \cite{CNQ15,MM08}.
The generation of zonal flows are caused by generation of small-scale
meridional Rossby waves by a modulational instability
and nonlinear interaction of the Rossby waves \cite{CNQ15}.
Generation of zonal flows in the Earth's atmosphere
occurs in unstably stratified flows.

Rossby waves also can be important in solar and stellar
astrophysics, e.g., in tachocline layer below the convection zone of the sun
and solar-like stars \cite{SZ92,GM98,ZO07}.
In plasma physics, Rossby waves are analogous to  the drift waves which
are caused by the combined effect of the drift motions perpendicular to the electric and magnetic fields and large-scale density gradient (see \cite{CNQ15,DII05},
and references therein).

It was suggested that Rossby waves can be related to the
El Ni\~{n}o and Southern Oscillation (see \cite{SS88,B88,MY99,VP03},
and references therein).
The El Ni\~{n}o is an irregularly periodical variation of winds and
sea surface temperatures over the tropical Pacific Ocean, while
the Southern Oscillation is an oscillation of surface air pressure
between the tropical eastern and the western Pacific Ocean.

\begin{figure}
\centering
\includegraphics[width=9.5cm]{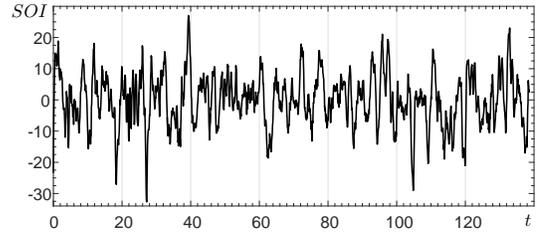}
\caption{\label{Fig1} Time dependence of the Southern Oscillation Index (SOI) after 5 month window averaging, where the time is measured in years, $t=0$ corresponds to 1878 year, the total time interval of the observations of the SOI is 138 years. The data are taken from \cite{DATA}.}
\end{figure}

\begin{figure}
\centering
\includegraphics[width=9.5cm]{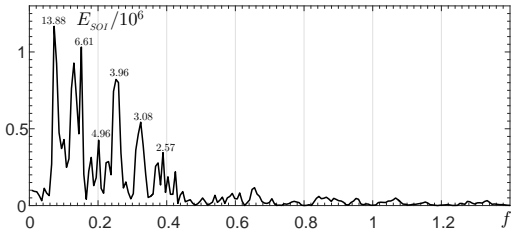}
\caption{\label{Fig2} The spectrum $E_{\rm SOI}(f)$ of the SOI. The frequency is measured in the units of the inverse years. The shown periods of oscillations (13.88; 6.61; 4.96; 3.96; 3.08; 2.57) are measured in years.}
\end{figure}

The strength of the Southern Oscillation is characterised by the Southern Oscillation
Index (SOI), and is determined from fluctuations of the surface air pressure
difference between Tahiti and Darwin, Australia
(see \cite{C14,P04}, and references therein).
For example in Fig.~\ref{Fig1} we show the time dependence of the SOI after 5 month window averaging, where the time is measured in years, $t=0$ corresponds to 1878 year,
and the total time interval of the observations of the SOI is 138 years.
Although this function looks like a random signal, the spectrum of the SOI
shown in Fig.~\ref{Fig2} contains several typical periods which are larger
than two years.

In spite of many observations of the SOI, it is not clear how Rossby waves
can explain these observations.
The classical two-dimensional Rossby waves are usually considered
in the $\beta$-plane approximation.
In the framework of this approximation, the motions are studied in
quasi two-dimensional incompressible thin fluid layer of constant density with a
free surface in hydrostatic balance on the rotating plane with variable
Coriolis parameter, that depends linearly on the horizontal coordinate (see \cite{P78,Z07},
and references therein).
However, the periods of the classical planetary Rossby waves investigated in numerous studies
do not exceed the time scales about 100 days.
On the other hand, the periods of oscillations related to the SOI, are larger than 2 years
[see Fig.~\ref{Fig2}].

The goal of the present study is to develop a theory of slow three-dimensional
Rossby waves in rotating spherical density stratified
convection, and apply the theory to explain the observations related to the Southern
Oscillation. This paper is organized as follows.
In Section~II, we consider the classical two-dimensional Rossby waves
in spherical geometry where we do not use the $\beta$-plane approximation.
In Section~III, we explain the basic ideas related to the theory of the
three-dimensional slow Rossby waves in rotating spherical density stratified
convection. In this section we also apply the developed theory for the
interpretation of the observed periods of the SOI.
In Section~IV, we discuss our results and draw some conclusions.
In Appendices A and B we present detailed
derivations related to the developed theory.

\section{Two-dimensional Rossby waves}

We start with traditional two-dimensional Rossby waves. To this end,
we neglect viscosity in Navier-Stokes equation for the velocity field
${\bm v}(t,{\bm r})$ and consider incompressible rotating flow:
\begin{eqnarray}
&& {\partial {\bm v} \over \partial t} + ({\bm v} {\bm \cdot} \bec{\nabla}){\bm v} = - {\bec{\nabla} p \over \rho} + 2 {\bm v} {\bm \times} {\bm \Omega} ,
\label{AA1}\\
&& \bec{\nabla} {\bf \cdot} {\bm v} =0 ,
\label{AA2}
\end{eqnarray}
where ${\bm \Omega}$ is the angular velocity, $p$ is the fluid pressure and $\rho$ is the fluid density. We use spherical geometry, $(r,\vartheta,\varphi)$ and consider the two-dimensional (2D) flow with $v_r=0$. This implies that we study the two-dimensional flow on the spherical surface with the constant radius $R$.
The equation for the radial component of vorticity $w_r(\vartheta,\varphi)$ is given by
\begin{eqnarray}
{\partial w_r \over \partial t} = - 2  ({\bm v} {\bm \cdot} \bec{\nabla}) {\bm \Omega}_r ,
\label{A1}
\end{eqnarray}
where ${\bm w}={\bm \nabla} {\bm \times} {\bm v}$ is the vorticity and ${\bm \Omega}=\Omega (\cos \vartheta, -\sin \vartheta, 0)$. Equation~(\ref{A1})
can be rewritten as follows:
\begin{eqnarray}
{\partial w_r \over \partial t} = {2 \Omega \over R}\,  v_{\vartheta} \, \sin\vartheta .
\label{A2}
\end{eqnarray}
For incompressible two-dimensional flow, we introduce the stream function $\Psi(\vartheta,\varphi)$ defined as:
\begin{eqnarray}
{\bm v} = \bec{\nabla} {\bm \times} \left[{\bm e}_r \, \Psi(\vartheta,\varphi) \, R\right] .
\label{A3}
\end{eqnarray}
Equations~(\ref{A2}) and~(\ref{A3}) yield the following equation for the stream function $\Psi(t,\vartheta,\varphi)$:
\begin{eqnarray}
{\partial \Delta \Psi \over \partial t} = -{2 \Omega \over R^2}\, {\partial \Psi \over \partial \varphi} ,
\label{A4}
\end{eqnarray}
where $w_r=-\Delta \Psi$ and $v_\vartheta= \sin^{-1} \vartheta \, (\partial \Psi / \partial \varphi)$.
We seek for a solution of this equation in the following form:
\begin{eqnarray}
\Psi(t,\vartheta,\varphi) =C_{\ell,m} \, P^{m}_\ell(\vartheta) \, \exp(i \omega t + im\varphi) ,
\label{A5}
\end{eqnarray}
where $P^{m}_\ell(\vartheta)$ is the associated Legendre function (spherical harmonics) of the first kind, that is the eigenfunction of the operator $R^2 \Delta$, i.e.,
\begin{eqnarray}
R^2 \Delta\, P^{m}_\ell(\vartheta) = -\ell(\ell+1) P^{m}_\ell(\vartheta) .
\label{A6}
\end{eqnarray}
Substituting Eq.~(\ref{A5}) into Eq.~(\ref{A4}) and using Eq.~(\ref{A6}),
we obtain the well-known expression for the frequency of the classical
two-dimensional Rossby waves in the incompressible flow \cite{P78}:
\begin{eqnarray}
\omega_{R}^{(2D)} = {2 \Omega m \over \ell(\ell+1)} .
\label{A7}
\end{eqnarray}

\section{Three-dimensional Rossby waves in convection}

In this section we study an excitation of slow three-dimensional $(3D)$ Rossby waves with frequency that is much smaller than the rotation frequency.
We consider a spherical rotating layer with the radius $R$ in the presence of a density stratified convection, use the anelastic approximation for very low-Mach-number flows, and
neglect dissipative terms, for simplicity.
The governing equations written in a rotating frame of reference read:
\begin{eqnarray}
&& {\partial {\bm v} \over \partial t} + ({\bm v} {\bm \cdot} \bec{\nabla}){\bm v} = -\bec{\nabla} \left({p \over \rho}\right) - \bec{\beta} \theta + 2 {\bm v} {\bm \times}
{\bm \Omega} ,
\label{B1}\\
&& {\partial \theta \over \partial t} + ({\bm v} {\bm \cdot} \bec{\nabla})\theta = -({\bm v} {\bf \cdot} \bec{\nabla})\theta_{\rm eq} ,
\label{B2}\\
&& \bec{\nabla} {\bf \cdot} (\rho \, {\bm v}) =0 ,
\label{B3}
\end{eqnarray}
where $\bec{\beta}= {\bm g}/T_0$ is the buoyancy parameter, ${\bm g}$ is the acceleration due to the gravity, $\theta=T \left(p_0/p\right)^{1-1/\gamma}$ is the potential temperature with equilibrium value $\theta_{\rm eq}$. Here $T$ is the physical temperature with the reference value $T_0$, $p$ is the fluid pressure with the reference value $p_0$ and $\rho$ is the fluid density, and $\gamma=c_{\rm p}/c_{\rm v}$ is the ratio of specific heats.

We linearize Eqs.~(\ref{B1})-(\ref{B3}), take twice ${\bf curl}$ from equation of motion~(\ref{B1}) to exclude pressure term, use spherical coordinates $r, \vartheta, \varphi$, and write the obtained equations for the radial components of velocity and vorticity $v_r, w_r$:
\begin{eqnarray}
&&\biggl(\bec{\Lambda} {\bm \cdot} \bec{\nabla}  - \Delta \biggr) {\partial v_{r} \over
\partial t} = 2 ({\bm \Omega} \cdot \bec{\nabla}) w_r + 2 ({\bm \Omega} {\bm \times}
{\bm \nabla})_r \, (\bec{\Lambda} {\bm \cdot} {\bm v})
\nonumber\\
&& \quad + 2 r^{-1} \, \Omega_\vartheta \, \nabla_r v_\varphi - \beta \Delta_{\perp} \theta ,
\label{B4}\\
&&{\partial w_r \over \partial t} = 2 ({\bm \Omega} \cdot \bec{\nabla}) v_r -
2 {\bm \Omega}_r (\bec{\Lambda} {\bm \cdot} {\bm v}) - 2  ({\bm v} {\bm \cdot} \bec{\nabla}) {\bm \Omega}_r,
\label{B5}\\
&&{\partial \theta \over \partial t} = -({\bm v} {\bm \cdot} \bec{\nabla})\theta_{\rm eq},
\label{B6}
\end{eqnarray}
where $\bec{\Lambda} =- \bec{\nabla}\rho/\rho =$ const, and the continuity equation~(\ref{B3}) reads $\bec{\nabla} {\bf \cdot} {\bm v} = \bec{\Lambda} {\bm \cdot} {\bm v}$. Let us
rewrite Eqs.~(\ref{B4})-(\ref{B6}) using new variables: ${\bm U}=\sqrt{\rho} \,{\bm v} $, ${\bm W}=\sqrt{\rho} \,{\bm w}$ and $\Theta=\sqrt{\rho} \,\theta$:
\begin{eqnarray}
&&\biggl({\Lambda^{2} \over 4} - \Delta \biggr) {\partial U_{r} \over
\partial t} = (2 {\bf \Omega} \cdot \bec{\nabla} +
{\bf \Omega} \cdot {\bm \Lambda}) W_r
\nonumber\\
& & \quad  + 2 ({\bm \Omega} {\bm \times}
{\bm \nabla})_r \, ({\bm \Lambda} {\bm \cdot} {\bm U})
+ r^{-1} \, \Omega_\theta \, \left(2\nabla_r + \Lambda\right)U_\varphi
- \beta \Delta_{\perp} \Theta ,
\nonumber\\
\label{B7} \\
&& {\partial W_r \over \partial t} = (2 {\bm \Omega} \cdot \bec{\nabla} -
{\bf \Omega} \cdot {\bm \Lambda}) U_{r} + {2 \Omega \over r}\,  U_{\vartheta} \, \sin\vartheta,
\label{B8} \\
&&{\partial \Theta \over \partial t} = - {\Omega_{b}^{2} \over \beta} U_{r} - U_\vartheta \nabla_\vartheta \Theta_{\rm eq},
\label{B9}\\
&&\bec{\nabla} {\bf \cdot} {\bm U} = {1 \over 2} \bec{\Lambda} {\bm \cdot} {\bm U} ,
\label{B32}
\end{eqnarray}
where $\Omega_{b}^{2}=\beta \nabla_r \Theta_{\rm eq}$ and $\Delta_{\perp}$ is the angular part of the Laplacian $\Delta$.
We seek for a solution of Eqs.~(\ref{B7})-(\ref{B32}) using a basis of spherical vector functions that are eigenfunctions of the operator ${\bf curl}$ in spherical coordinates.
In particular, ${\bm U}$ is given by the following expression:
\begin{eqnarray}
{\bm U} &=& \sum_{\ell,m} \, \exp\left(\lambda t - ik \, r\right)
\biggl[\hat A_{\rm r} {\bm Y}_{\rm r} + \hat A_{\rm p} {\bm Y}_{\rm p} + \hat A_{\rm t} {\bm Y}_{\rm t} \biggr],
\nonumber\\
\label{B12}
\end{eqnarray}
where the coefficients $\hat A_{\rm r}, \hat A_{\rm p}, \hat A_{\rm t}$ depend on $\ell$ and $m$, and the vector basis functions (radial, poloidal and toroidal) are given by
\begin{eqnarray}
&& {\bm Y}_{\rm r}={\bm e}_r Y_\ell^m(\vartheta,\varphi), \quad {\bm Y}_{\rm p} =r \,\bec{\nabla} Y_\ell^m(\vartheta,\varphi),
\nonumber\\
&& {\bm Y}_{\rm t} = ({\bm r} {\bm \times}{\bm \nabla})
Y_\ell^m(\vartheta,\varphi),
\label{BB12}
\end{eqnarray}
and $Y_\ell^m(\vartheta,\varphi)$ is the eigenfunction of the operator $r^2 \, \Delta_\perp$
with the eigenvalue $-\ell(\ell+1)$.

Using the procedure described in Appendix~A we obtain the following
dispersion equation:
\begin{eqnarray}
&& \left(\lambda - 2 i \omega_{R}^{(2D)} \right) \, \left(\lambda + i \omega_{R}^{(3D)} \right) + 2 \Omega^2 \left(1 + {3 i \sigma \over \ell^2 \,(1 + \sigma^2)} \right)
\nonumber\\
&& \quad +  {\Omega_{b}^{2} H_\rho \over R} \, \left({i \omega_{R}^{(3D)} \over \lambda}
+ {4 \ell^2 \, H_\rho \over R \,(1 + \sigma^2)} \right)
\nonumber\\
&& \quad - \omega_{R}^{(2D)} \left(\omega_{R}^{(2D)} + \omega_{R}^{(3D)}\right)=0 ,
\label{BBB43}
\end{eqnarray}
where $\ell^2 \gg 1$, $\sigma=2k \, H_\rho$, $H_\rho=1/|{\bm \Lambda}|$,
\begin{eqnarray}
\omega_{R}^{(3D)} &=& {8 m \,\Omega \, H_\rho \over R \,(1 + \sigma^2)},
\label{BB13}
\end{eqnarray}
and $\omega_{R}^{(2D)}$ is the frequency of the two-dimensional Rossby waves determined by Eq.~(\ref{A7}).
Equation~(\ref{BBB43}) determines three modes: the convective mode,
the inertial waves and the three-dimensional Rossby waves.
These Rossby waves are due to the combined effect of
rotation and curvature of the surface in density stratified
developed convection.
The inertial waves are caused only by rotation and described by the following dispersion relation: $\omega_I = 2(\bec{\Omega} \cdot {\bm k})/ k$, where ${\bm k}$
is the wave vector.

Let us justify the assumption behind the condition $\ell^2 \gg
1; m^2$ used for the derivation of Eqs.~(\ref{BBB43}) and~(\ref{BB13}).
We study excitation of the three-dimensional slow Rossby waves from
the equilibrium based on the developed convection.
According to the observations (see, e.g., \cite{V06}),
6 large-scale convective cells are observed in the Earth's
atmospheric  large-scale circulation (the Hadley cell, the Mid-latitude cell and the Polar cell, i.e., there are 3 convective cells in the Northern Hemisphere and 3 convective
cells in the Southern Hemisphere).
These cells can be interpreted in terms of the solution
of Eqs.~(\ref{B7})-(\ref{B32}) in the form of the spherical
functions $Y_\ell^m$ with $\ell=6$ and $m=0$
(i.e., the axisymmetric modes).
The  solutions for the slow three-dimensional Rossby waves
are not axisymmetric, and they have $m=1$ or $m=2$.
This is the reason for the assumption that $\ell(\ell+1) \gg 1; m^2$.

Let us determine the conditions for the existence of the slow 3D Rossby waves.
The balance $2 \Omega^2 + (\Omega_{b}^{2} H_\rho / R) \, (i \omega_{R}^{(3D)} / \lambda) =0$ with $\lambda = - i \omega_{R}^{(3D)}$ determines the
Brunt-V\"{a}is\"{a}l\"{a} frequency:
\begin{eqnarray}
\Omega_{b} = \Omega \, \left({2 \, R\over H_\rho}\right)^{1/2},
\label{B45}
\end{eqnarray}
so the frequency of the three-dimensional slow Rossby waves is given by:
\begin{eqnarray}
\omega_{R} = - \omega_{R}^{(3D)} .
\label{B13}
\end{eqnarray}
The frequencies of the slow three-dimensional Rossby waves, $\omega_{R}^{(3D)}$,
are smaller than some typical frequencies in the system:
$\omega_{R}^{(3D)} \ll \omega_{R}^{(2D)} \ll \Omega \ll |\Omega_{b}|$.
In particular, since $\ell \gg m$ the frequency of the 2D Rossby waves $\omega_{R}^{(2D)}$ is much smaller than the angular velocity $\Omega$ [see Eq.~(\ref{A7})].
Since $H_\rho \ll R$, the frequency of the 3D Rossby waves $\omega_{R}^{(3D)}$ is much less than the angular velocity $\Omega$.
The period of rotation, $T_\Omega=2 \pi /\Omega$ is 24 hours,
the periods of the classical planetary two-dimensional
Rossby waves $T_{R}^{(2D)} = 2 \pi /\omega_{R}^{(2D)}$ do not exceed
the time scales of the order of 100 days.
In the Earth's atmosphere, the periods of the slow three-dimensional Rossby waves,
$T_R=2 \pi / \omega_{R}^{(3D)}$, vary in the range from 3 to 14 years (see below).

In Fig.~\ref{Fig3} we show the period $T_R =2 \pi / |\omega_{R}^{(3D)}|$ of the slow Rossby waves as a function of the normalized characteristic scale $L_z/H_\rho$, where $L_z = \pi /k$. Let us invoke the three-dimensional slow Rossby waves for the interpretation of the observed periods of the Southern Oscillation.
The Southern Oscillation is characterised by the Southern Oscillation
Index (SOI) (see Fig.~\ref{Fig2}), that contains several typical periods which are larger
than two years. For example, the period $T_R =2 \pi / |\omega_{R}^{(3D)}|= 13.88$ years is obtained for $L_z= 7.3$ km, where in Eqs.~(\ref{BB13}) and~(\ref{B13}) we took into account that $m=1$, $H_\rho=8.3$ km, $\Omega=2.27 \times 10^3$
rad/years and $R=6400$ km. Here $k=\pi/L_z$ corresponds to the boundary conditions:
$v_r(r=R)=v_r(r=R+L_z)=0$ and $v_\vartheta(r=R)\approx -v_\vartheta(r=R+L_z)$.
Similarly, for $m=1$, the period equals $T_R =6.61$ years for $L_z= 10.6$ km; $T_R =4.96$ years is obtained for $L_z= 12.4$ km and $T_R =3.96$ years is obtained for $L_z= 16.3$ km.
On the other hand, for $m=2$, we obtain that the period equals $T_R =3.96$ years for $L_z= 9.69$ km and $T_R =2.57$ years is obtained for $L_z= 12.1$ km.
Note that the vertical scales $L_z$ are of the order of the height of the tropopause.
These estimates and Fig.~\ref{Fig3} show that the three-dimensional slow Rossby waves are able to describe the observed periods of the SOI.

\begin{figure}
\centering
\includegraphics[width=9.0cm]{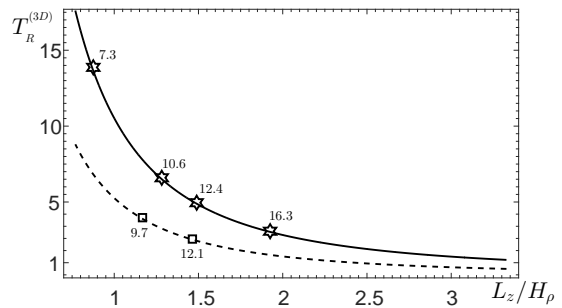}
\caption{\label{Fig3} The period $T_R =2 \pi / |\omega_{R}^{(3D)}|$ (measured in years) of the slow 3D Rossby waves versus the characteristic scale $L_z/H_\rho$, where $L_z = \pi /k$ for $m=1$ (solid) and $m=2$ (dashed).
The stars (for $m=1$) and squares (for $m=2$) correspond to the observed periods of the SOI, see Fig.~\ref{Fig2}. The vertical sizes of these observed modes, $L_z= 7.3; 10.6; 12.4; 16.3$ km, determine the wave numbers $k$ of the slow 3D Rossby waves, see Eqs.~(\ref{BB13}) and~(\ref{B13}).}
\end{figure}

Note that the frequency of the classical Rossby waves,
\begin{eqnarray}
\omega = -{\beta_f k_x \over k_x^2 + k_y^2 + R_d^{-2}},
\label{D100}
\end{eqnarray}
is obtained, e.g., using the $\beta$-plane approximation and assuming that
the Coriolis parameter $f=f_0 + \beta_f y$
(see \cite{P78,Z07}, and references therein).
Here $f_0=2 \Omega \sin \phi$, $R_d=\sqrt{g H} /f_0$ is the Rossby deformation radius,
$H$ is the average depth of the layer, $\phi$ is the contact point latitude,
the coordinate $y$ is in the meridional direction (northward in the geophysical context),
the coordinate $x$  is in the zonal direction (eastward),
and the vertical coordinate $z$ is in the direction normal
to the tangent plane, i.e. opposite to the effective gravity ${\bm g}$,
which includes a small correction due to the centrifugal acceleration.
When $\beta_f$ is small (i.e., the slope of a rotating
container is sufficiently small), the frequency of the Rossby waves is small,
i.e., the period of the Rossby waves is large. However, when Eq.~(\ref{D100})
is applied to the Earth's atmosphere, it does not yield the time scales
ranging from 3 to 14 years (which are the typical time scales for the SOI).
In particular, the periods of the classical Rossby waves described by
Eq.~(\ref{D100}), do not exceed the time scales of the order of 100 days
in the Earth's atmosphere.

\begin{figure}
\centering
\includegraphics[width=9.0cm]{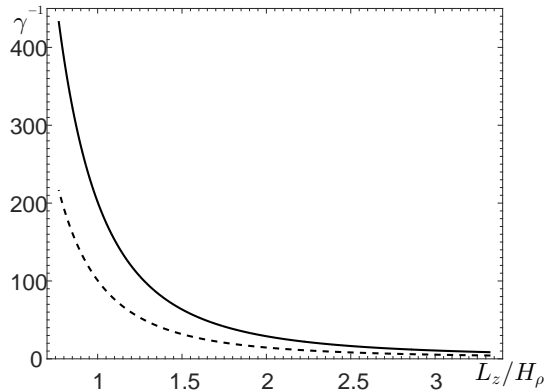}
\caption{\label{Fig4} The characteristic time $T_{\rm inst} = \gamma_{R}^{-1}$ (measured in years) of the excitation of the 3D Rossby waves versus the characteristic scale $L_z/H_\rho$ of the convective cells for $m=1$ (solid) and $m=2$ (dashed).}
\end{figure}

Let us discuss the mechanism of excitation of the slow Rossby waves.
The dispersion equation~(\ref{BBB43}) allows us to determine
the growth rate of the instability that excites the slow Rossby waves:
\begin{eqnarray}
\gamma_{R} = {3 \sigma \omega_{R}^{(3D)}\over \ell^2 (1 + \sigma^2)} .
\label{B14}
\end{eqnarray}
In Fig.~\ref{Fig4} we show the characteristic time $T_{\rm inst} = \gamma_{R}^{-1}$ of the excitation of the three-dimensional slow Rossby waves versus the characteristic scale $L_z/H_\rho$.
This instability is a non-axisymmetric instability modified by rotation
in a density stratified convection.
The instability causes excitation of the 3D Rossby waves interacting
with the convective mode and the inertial wave mode.
The energy of this instability
is supplied by thermal energy of convective motions.
This instability is different from the convective instability
in unstably stratified flows. In particular, the convective instability
can be excited in the axisymmetric flows.

\section{Discussion}

In the present study we propose a theory of three-dimensional slow Rossby waves in rotating spherical density stratified convection.
The frequency of these waves [see Eq.~(\ref{BB13})] is approximately determined by the following balance in Eq.~(\ref{B7}):
\begin{eqnarray}
&&\biggl({\Lambda^{2} \over 4} - \Delta \biggr) {\partial U_{r} \over
\partial t} \sim  2 ({\bm \Omega} {\bm \times} {\bm \nabla})_r
\, (\bec{\Lambda} {\bm \cdot} {\bm U}).
\label{D1}
\end{eqnarray}
This implies that the slow Rossby waves exist only in three-dimensional
density stratified flows, while in the incompressible flows there
occur classical two-dimensional Rossby waves (with vanishing radial velocity)
which exist even without convection.
The slow three-dimensional Rossby waves emerge when the convective mode is
approximately balanced by the inertial mode, i.e., there is a balance between
the Coriolis and buoyancy forces.
The slow Rossby waves are excited by a non-axisymmetric instability
with the characteristic time that is much larger than the period of these waves.
In this study we have not taken into account the effects of turbulence, e.g.,
turbulence causes dissipation of waves. In atmospheric flows, the strongest turbulence is located at the turbulent boundary layer at the height that is of the order of 1 km.

We apply the developed theory for the three-dimensional slow Rossby waves
to interpret the observations of the Southern Oscillation
characterized by the Southern Oscillation Index (SOI).
We found a good agreement between the theoretical results
and observations of the SOI.

\begin{acknowledgements}
This research was supported in part by the Israel
Science Foundation governed by the Israeli
Academy of Sciences (Grant No. 1210/15).
\end{acknowledgements}

\appendix

\section{Derivation of the dispersion equation~(\ref{BBB43})}

We seek for a solution of Eqs.~(\ref{B7})-(\ref{B32}) using a basis of spherical vector functions that are the eigenfunctions of the operator ${\bf curl}$ in spherical coordinates.
The function ${\bm U}$ is given by
${\bm U} = \sum_{\ell,m} \, \exp(\lambda t - ik \, r)
[\hat A_{\rm r} {\bm Y}_{\rm r} + \hat A_{\rm p} {\bm Y}_{\rm p} + \hat A_{\rm t} {\bm Y}_{\rm t}]$, where the coefficients $\hat A_{\rm r}, \hat A_{\rm p}, \hat A_{\rm t}$ depend on $\ell$ and $m$, and
${\bm Y}_{\rm r}={\bm e}_r Y_\ell^m(\vartheta,\varphi)$, $\, {\bm Y}_{\rm p} =r \,\bec{\nabla} Y_\ell^m(\vartheta,\varphi)$,
$\, {\bm Y}_{\rm t} = ({\bm r} {\bm \times}{\bm \nabla})Y_\ell^m(\vartheta,\varphi)$,
and $Y_\ell^m(\vartheta,\varphi)$. The properties of these functions are given in Appendix B.
Using Eqs.~(\ref{ABB34})--(\ref{ABB35}) given in Appendix B, we obtain:
\begin{eqnarray}
W_r &=& - r^{-1} \sum_{\ell,m} \ell(\ell+1) \, \hat A_{\rm t} \,({\bm e}_r \cdot {\bm Y}_{\rm r}) \, \exp\left(\lambda t - ik \, r\right) .
\nonumber\\
\label{B16}
\end{eqnarray}
Using Eqs.~(\ref{B12}), (\ref{B16}) and~(\ref{B21}) we get:
\begin{eqnarray}
U_{\vartheta} = \left[\left({\Lambda \over 2} -\nabla_r - {2 \over r} \right) \nabla_{\vartheta} \Delta_\perp^{-1}\, U_{r} - \nabla_\varphi \Delta_\perp^{-1}\, W_r \right],
\nonumber\\
\label{B36}\\
U_{\varphi} =  \left[\left({\Lambda \over 2} - \nabla_r - {2 \over r} \right) \nabla_{\varphi} \Delta_\perp^{-1} \, U_{r}+ \nabla_\vartheta \Delta_\perp^{-1} \,
W_r \right] .
\nonumber\\
\label{B37}
\end{eqnarray}
Using Eqs.~(\ref{B36}) and~(\ref{B37}), we rewrite Eqs.~(\ref{B7})-(\ref{B8}) in the following form:
\begin{eqnarray}
\hat{\cal L} W_r &=& \hat\Omega^{(-)} U_{r} ,
\label{NB7}\\
\hat{\cal M} U_r &=& \hat\Omega^{(+)} W_{r} - \beta \Delta_{\perp} \Theta ,
\label{NB8}
\end{eqnarray}
where
\begin{eqnarray}
&&\hat{\cal M} = \biggl({\Lambda^{2} \over 4} - \Delta \biggr) {\partial \over
\partial t} + {2 \Omega \over r^2} \biggl[\left({\Lambda^2 \over 4} - \nabla_r^2\right) \Delta_{\perp}^{-1} + \Lambda \, r \biggr] {\partial \over \partial \varphi} ,
\nonumber\\
\label{B40}\\
&&\hat{\cal L} = {\partial \over \partial t} + {2 \Omega \over r^2} {\partial \over \partial \varphi} \, \Delta_{\perp}^{-1} ,
\label{B41}\\
&& \hat\Omega^{(\pm)} = 2 \left({\bf \Omega} \cdot \bec{\nabla}\right) \pm
\biggl[- {2 \Omega \sin \vartheta \over r} \biggl(\nabla_r
+ {\Lambda \over 2}\biggr) \nabla_\vartheta \Delta_{\perp}^{-1}
\nonumber\\
&& \quad + {\bf \Omega} \cdot {\bf \Lambda} \biggr] .
\label{NB42}
\end{eqnarray}
The system of Eqs.~(\ref{NB7})-(\ref{NB8}) is reduced to the following equation:
\begin{eqnarray}
&& \biggl[\left(\hat{\cal L} \hat{\cal M} - \Omega_{b}^{2} \, \Delta_{\perp} \, \hat H - \hat\Omega^{(+)} \, \hat\Omega^{(-)} \right) {\partial \over \partial t}
\nonumber\\
&& \quad \quad  - \Omega_{b}^{2} \, \hat H \, {2 \Omega \over r^2} {\partial \over \partial \varphi} - \Omega_{b\vartheta}^2 \, \hat\Omega^{(-)} \, \nabla_\varphi \biggr] \zeta_r =0 ,
\label{B10}
\end{eqnarray}
where $U_r = \partial \zeta_r / \partial t$,
\begin{eqnarray}
\hat{\cal H} = 1 + \left({\Omega_{b\vartheta}^2 \over  \Omega_{b}^2} \right) \left(-\nabla_r + {\Lambda \over 2}\right) \nabla_\vartheta \, \Delta_\perp^{-1},
\label{BB42}
\end{eqnarray}
and $\Omega_{b\vartheta}^{2}=\beta \, ({\bm e}_\vartheta \cdot {\bm \nabla}_\vartheta)
\, \Theta_{\rm eq}$. Using Eqs.~(\ref{B10}), we arrive at the
dispersion equation~(\ref{BBB43}) for $\ell^2 \gg 1$.

Using Eqs.~(\ref{B36})--(\ref{NB7}), we determine the components of the function ${\bm U}$
and the radial component of the function ${\bm W}$
for the three-dimensional slow Rossby waves:
\begin{eqnarray}
U_r &=& i V_0 \, \exp(\gamma_{R} t) \, \exp\left[i(\omega_{R} t - k \, r)\right] \, Y_\ell^m,
\label{C39}\\
U_\theta &=& {R \, V_0 \, \exp(\gamma_{R} t) \over 4 H_\rho \, \sin \theta} \, \exp\left[i(\omega_{R} t - k \, r)\right] \, (i -\sigma)
\nonumber\\
&& \times  \Big(Y_{\ell-1}^m + Y_{\ell+1}^m \Big) ,
\label{C40}\\
U_\varphi &=& {R \, V_0 \, \exp(\gamma_{R} t) \over 8 m \, H_\rho \, \sin \theta} \,
\exp\left[i(\omega_{R} t - k \, r)\right] \, \biggl\{(3i\sigma - 1)\, Y_{\ell}^m
\nonumber\\
&& + \Big[i \sigma \, (\ell-1) + \ell+1 \Big] \, Y_{\ell-2}^m - \Big[i \sigma \, (\ell+2) + \ell \Big] \, Y_{\ell+2}^m \biggr\} ,
\nonumber\\
\label{C41}\\
W_r &=& {V_0 \, \exp(\gamma_{R} t) \over 4 m \, H_\rho} \exp\left[i(\omega_{R} t - k \, r)\right] \,  \biggl\{(\ell-1) \, \Big[i \sigma \, (\ell-1)
\nonumber\\
&& + \ell+1 \Big] \, Y_{\ell-1}^m + (\ell+2) \, \Big[i \sigma \, (\ell+2) + \ell \Big] \, Y_{\ell+1}^m \biggr\}  ,
\nonumber\\
\label{C42}
\end{eqnarray}
where $V_0 = - i \hat A_{\rm r}$.
The equation for the function $P=\sqrt{\rho} \, \,p$ is obtained by calculating the divergence of Eq.~(\ref{B1}) and using Eq.~(\ref{C42}):
\begin{eqnarray}
P &=& - {\rho V_0 \, \Omega \, H_\rho \, \exp(\gamma_{R} t) \over m \, (\sigma + i)^2}
\exp\left[i(\omega_{R} t - k \, r)\right]
\nonumber\\
&& \times \biggl\{(\ell-1) \, \Big[i \sigma \, (\ell-1)+ \ell+1 \Big] \, \Big(Y_{\ell-2}^m + Y_{\ell}^m \Big)
\nonumber\\
&& + (\ell+2) \, \Big[i \sigma \, (\ell+2) + \ell \Big] \, \Big(Y_{\ell+2}^m + Y_{\ell}^m \Big) \biggr\}  .
\label{C43}
\end{eqnarray}

\section{Properties of the spherical vector functions}

The spherical vector functions are the eigenfunctions of the operator ${\bf curl}$ in spherical coordinates. These functions can be presented as
\begin{eqnarray}
Y_\ell^m(\vartheta,\varphi) =A_{\ell,m} \, P^{|m|}_\ell(\vartheta) \, \exp(im\varphi) ,
\label{B30}
\end{eqnarray}
where the coefficient $A_{\ell,m}$ is determined by the normalization condition,
$\int Y_\ell^m(\vartheta,\varphi) \, Y_n^{-m}(\vartheta,\varphi) \, \sin \vartheta \,d\vartheta \,d\varphi =\delta_{\ell n}$, and is given by
\begin{eqnarray}
A_{\ell,m} = {1 \over 2} \left[{(\ell-|m|)! \over (\ell+|m|)!} \, \left({2\ell+1 \over \pi}\right)\right]^{1/2} .
\label{B31}
\end{eqnarray}
(see \cite{AS72,N82}, and references therein).
Equation~(\ref{B31}) yields the following identity:
\begin{eqnarray}
{A_{\ell,m} \over A_{\ell+1,m}} \left[{\ell + 1 -|m|
\over 2\ell + 1} \right]
= \left[{(\ell + 1)^2 -m^2 \over 4(\ell + 1)^2 -1}\right]^{1/2} .
\nonumber\\
\label{C38}
\end{eqnarray}
The associated Legendre function (spherical harmonics) of the first kind, $P^{m}_\ell(\vartheta)$, is determined by the following equation:
\begin{eqnarray}
&& \left[(1-Z^2) {d^2 \over dZ^2} - 2Z {d \over dZ} + \left(\ell(\ell+1) - {m^2 \over 1-Z^2} \right)\right] P^{m}_\ell(Z)
\nonumber\\
&& \quad =0,
\label{BB21}
\end{eqnarray}
where $Z=\cos \vartheta$.
The function $P^{m}_\ell(Z)$ has the following properties:
\begin{eqnarray}
&&(2\ell+1) Z P_\ell^m(Z) = (l-m+1) P^{m}_{\ell+1}(Z)
\nonumber\\
&&\quad + (\ell+m) P^{m}_{\ell-1}(Z),
\label{BB19}\\
&& Z^2 P_\ell^m(Z) = I_{\ell,m} P_\ell^m(Z) + J_{\ell,m} P_{\ell+2}^m(Z)
+L_{\ell,m} P_{\ell-2}^m(Z) ,
\nonumber\\
\label{B19}
\end{eqnarray}
where
\begin{eqnarray}
I_{\ell,m} = {\ell^2-m^2 \over 4 \ell^2 -1} + {(\ell+1)^2-m^2 \over 4 (\ell+1)^2 -1} .
\label{B20}
\end{eqnarray}
When $\ell^2 \gg 1; m^2$, the function $I_{\ell,m}\approx 1 / 2$.
We also took into account that
\begin{eqnarray}
&&\int_0^{\pi} \sin^2 \vartheta \cos \vartheta \, P_\ell^m(\vartheta) \, {\partial \over \partial\vartheta} P_\ell^{m}(\vartheta) \,d\vartheta
\nonumber\\
&& \quad = {1 \over 2} \, \int_{-1}^{1} (Z^3 -Z) \, {\partial \over \partial Z} [P_\ell^{m}(Z)]^2 \,dZ = - {1 \over 4},
\label{BB20}\\
&&\int_0^{\pi} \sin^3 \vartheta \, P_\ell^m(\vartheta) \, {\partial^2 \over \partial\vartheta^2} P_\ell^{m}(\vartheta) \,d\vartheta
\nonumber\\
&& \quad = {1 \over 2} \left[2 m^2 - \ell(\ell+1) -1\right] .
\label{BB21}
\end{eqnarray}

Using Eqs.~(\ref{C38}) and~(\ref{BB19}) we obtain the following identities
for $\ell^2 \gg 1; m^2$:
\begin{eqnarray}
&& Z Y_\ell^m = \left[{\ell^2 -m^2 \over 4\ell^2 -1}\right]^{1/2} \, Y_{\ell-1}^m
+ \left[{(\ell + 1)^2 -m^2 \over 4(\ell + 1)^2 -1}\right]^{1/2} \, Y_{\ell+1}^m
\nonumber\\
&& \quad \approx {1 \over 2} \Big(Y_{\ell-1}^m + Y_{\ell+1}^m\Big),
\label{C36}\\
&& (Z^2 - 1) {dY_\ell^m \over dZ} \equiv r \, \sin \vartheta \, \nabla_\vartheta \, Y_\ell^m= \left[{(\ell + 1)^2 -m^2 \over 4(\ell + 1)^2 -1}\right]^{1/2}
\nonumber\\
&& \quad \times \ell \, Y_{\ell+1}^m - \left[{\ell^2 -m^2 \over 4\ell^2 -1}\right]^{1/2} \, (\ell + 1) \, Y_{\ell-1}^m
\nonumber\\
&& \quad \approx {1 \over 2} \Big(\ell \, Y_{\ell+1}^m - (\ell + 1) \, Y_{\ell-1}^m\Big),
\label{C37}\\
&& Z^2 Y_\ell^m \approx {1 \over 4} \Big(Y_{\ell-2}^m + 2Y_{\ell}^m + Y_{\ell+2}^m\Big),
\label{CC35}\\
&& \left(\sin \vartheta \, \nabla_\vartheta \, \Delta_{\perp}^{-1}\right) \, Y_\ell^m \approx {r \over 2\ell} \,\left(Y_\ell^m- {\ell \over \ell+1} \, Y_{\ell+1}^m \right),
\label{CC36}\\
&& \left(\sin \vartheta \, \nabla_\vartheta \, \Delta_{\perp}^{-1}\right) \, \left(\sin \vartheta \, \nabla_\vartheta \, \Delta_{\perp}^{-1}\right)  Y_\ell^m \approx {r^2 \over 4} \, \biggl[{Y_{\ell-2}^m \over \ell(\ell-1)}
\nonumber\\
&&  \quad + {Y_{\ell+2}^m \over (\ell+1)(\ell+2)}  - Y_\ell^m \, \left[\ell^{-2} + (\ell+1)^{-2} \right] \biggr] .
\label{CC37}
\end{eqnarray}
Since $\bec{\nabla} {\bf \cdot} (\rho \, {\bm v}) =0$, the coefficient $\hat A_{\rm p}$ is related with the coefficient $\hat A_{\rm r}$:
\begin{eqnarray}
\hat A_{\rm p}=- {ikr - 2 + r \Lambda/2 \over \ell(\ell+1)} \hat A_{\rm r} ,
\label{B21}
\end{eqnarray}
where we used the following properties of the spherical functions:
\begin{eqnarray}
\bec{\nabla} {\bm \cdot} {\bm Y}_{\rm r} &=& {2 \over r} \,({\bm e}_r \cdot {\bm Y}_{\rm r}),
\quad \bec{\nabla} {\bm \cdot} {\bm Y}_{\rm p} =-{\ell(\ell+1) \over r} \,({\bm e}_r \cdot {\bm Y}_{\rm r}),
\nonumber\\
\bec{\nabla} {\bm \cdot} {\bm Y}_{\rm t} &=& 0 .
\label{B33}
\end{eqnarray}
To derive Eq.~(\ref{B16}) we took into account that:
\begin{eqnarray}
\bec{\nabla} {\bm \times} {\bm Y}_{\rm r} &=& - \bec{\nabla} {\bm \times} {\bm Y}_{\rm p} = - {1 \over r} \, {\bm Y}_{\rm t} ,
\nonumber\\
\bec{\nabla} {\bm \times} {\bm Y}_{\rm t} &=& - {1 \over r} \left[\ell(\ell+1)\, {\bm Y}_{\rm r}+ {\bm Y}_{\rm p}\right] ,
\label{ABB34}\\
{\bm Y}_{\rm p} &=& - {\bm e}_r {\bm \times} {\bm Y}_{\rm t} , \quad {\bm Y}_{\rm t} = {\bm e}_r {\bm \times} {\bm Y}_{\rm p} .
\label{ABB35}
\end{eqnarray}
Using Eqs.~(\ref{B33})-(\ref{ABB35}) we obtain:
\begin{eqnarray}
\bec{\nabla} {\bm \cdot} {\bm U} &=& -{1 \over r} \sum_{\ell,m} \exp\left(\lambda_{\ell,m} t - ik \, r\right) \biggl[(i k r -2)\hat A_{\rm r}\,
\nonumber\\
&& + \ell(\ell+1) \, \hat A_{\rm p} \biggr]\,({\bm e}_r \cdot {\bm Y}_{\rm r}) \,,
\label{B17}\\
\bec{\nabla} {\bm \times} {\bm U} &=& -{1 \over r} \sum_{\ell,m} \, \exp\left(\lambda_{\ell,m} t - ik \, r\right) \biggl[ \left(\hat A_{\rm r}\, - i k r\, \hat A_{\rm p} \right) {\bm Y}_{\rm t}
\nonumber\\
&& + (i k r +1) \, \hat A_{\rm t} {\bm Y}_{\rm p} + \ell(\ell+1) \, r^{-1} \, \hat A_{\rm t} {\bm Y}_{\rm r} \biggr] .
\label{B18}
\end{eqnarray}

\end{document}